\documentclass[aps,prl,floatfix,twocolumn,reprint,amsmath,amssymb,superscriptaddress]{revtex4-2}

\usepackage{amsfonts}
\usepackage{mathrsfs}
\usepackage{amsmath}
\usepackage{color}
\usepackage{graphicx}
\usepackage{bm}
\usepackage{amssymb}
\usepackage{xspace}
\usepackage{epstopdf}
\usepackage{dcolumn}
\usepackage{longtable}
\usepackage{multirow}
\usepackage{float}
\usepackage{comment}
\usepackage{soul}
\usepackage[colorlinks=true, linkcolor=blue, citecolor=blue, anchorcolor=blue]{hyperref}

\makeatother

\begin{document}

\title{A General Theory of Chiral Splitting of Magnons in Two-Dimensional Magnets}

\author{Yu Xie}
\email{These authors contributed equally to this work.}
\affiliation{School of Materials Science and Physics, China University of Mining and Technology, Xuzhou 221116, China}
\author{Dinghui Wang}
\email{These authors contributed equally to this work.}
\affiliation{School of Materials Science and Physics, China University of Mining and Technology, Xuzhou 221116, China}
\affiliation{Jiangsu Physical Science Research Center, Nanjing 210093, China}
\author{Chao Li}
\affiliation{School of Materials Science and Physics, China University of Mining and Technology, Xuzhou 221116, China}
\author{Xiaofan Shen}
\affiliation{School of Materials Science and Physics, China University of Mining and Technology, Xuzhou 221116, China}
\author{Junting Zhang}
\email{juntingzhang@cumt.edu.cn}
\affiliation{School of Materials Science and Physics, China University of Mining and Technology, Xuzhou 221116, China}

\date{\today}

\begin{abstract}
Magnons in antiferromagnets exhibit two chiral modes, providing an intrinsic degree of freedom for magnon-based computing architectures and spintronic devices. Electrical control of chiral splitting is crucial for applications, but remains challenging. Here, we propose the concept of extrinsic chiral splitting, involving alternating and ferrimagnet-like types, which can be induced and controlled by an electric field. A symmetry framework based on 464 collinear spin layer groups is established to classify chiral splitting characteristics and electric field responses in two-dimensional magnets. We further elucidate how the spin layer group determines the type of alternating chiral splitting and the dominant lowest-order magnetic exchange interaction. We demonstrate electric-field control over the magnitude and sign of the chiral splitting, enabling control of the spin Seebeck and Nernst effects related to thermal spin transport. This work provides a general theory for electric field manipulation of magnon chirality, paving the way for low-power magnonic logic devices.
\end{abstract}

\maketitle
Magnons, as quasiparticles of spin-wave excitations, carry spin angular momentum without charge, enabling dissipationless spin transport \cite{Chumak2015, Kruglyak2010, Pirro2021}. Furthermore, the scalability down to the atomic scale and nanometer wavelengths make them promising for applications in low-power, high-density logic circuits \cite{Khitun2010, Yu2014} and neuromorphic computing \cite{Grollier2020}. Currently, the manipulation of magnon transport has been explored in some typical ferromagnets (FM) and ferrimagnets (FiM), such as Yttrium Iron Garnet thin films, where spin current can be modulated via spin-orbit torque effect \cite{Cornelissen2015, Wang2024}. Compared to FM and FiM systems, magnons in antiferromagnets (AFM) exhibit ultrafast dynamics and exceptional robustness against stray magnetic field \cite{Jungwirth2016, Chumak2015, Kruglyak2010, Pirro2021}. In AFM, magnons exhibit both left-handed and right-handed chirality, which provides an intrinsic degree of freedom for magnon-chirality-based computing architectures and spintronic devices \cite{Jungwirth2016, Cheng2014, Daniels2015}. However, AFM typically exhibit chiral degeneracy of magnons, so breaking this chiral degeneracy is a prerequisite for controlling spin transport. Experimentally, external magnetic fields have been employed to lift chiral degeneracy, thereby inducing observable spin transport phenomena \cite{Lebrun2018, Seki2015, Li2019}.

\begin{figure*}
	\centering
	\includegraphics[width=2\columnwidth]{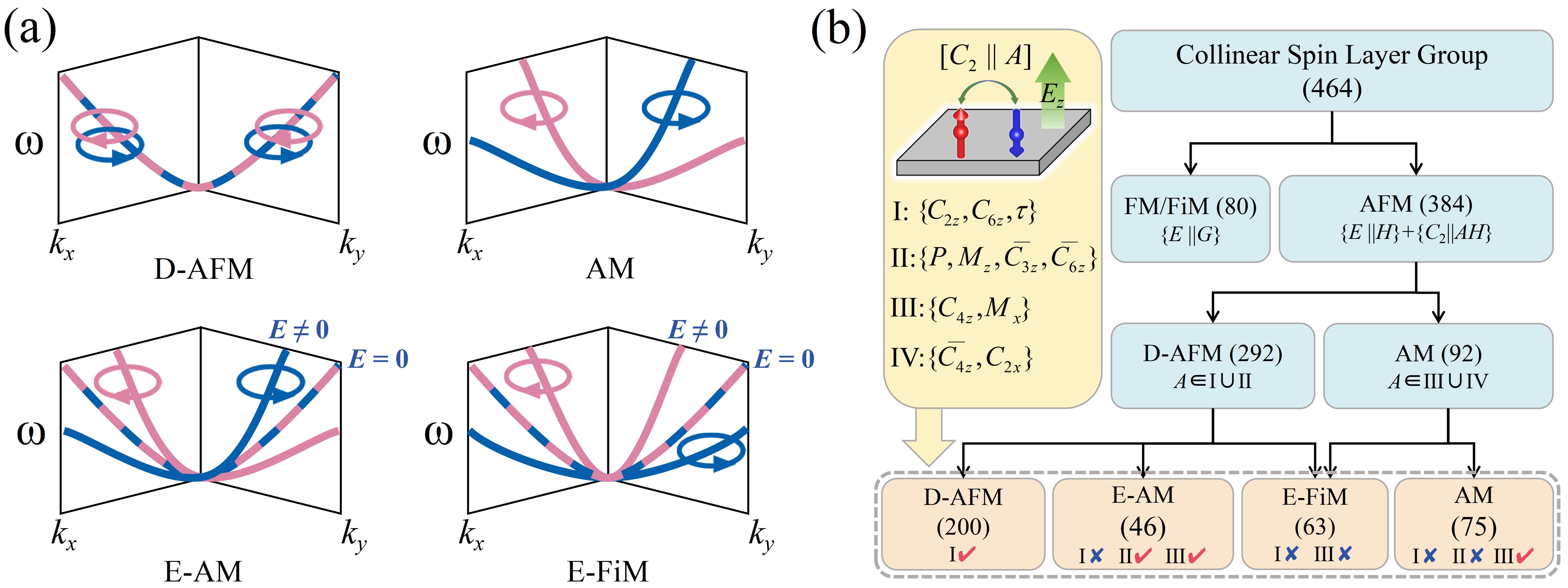}
	\caption{Characteristics of chiral splitting of magnon bands and symmetry classification of electric-field responses of 2D collinear magnets. (a) Schematic magnon band structures for degenerate AFM (D-AFM), AM, Extrinsic AM (E-AM) and FiM (E-FiM). The two magnon chiralities are distinguished by different colors. (b) Classification of the collinear spin layer groups (SLGs). The SLGs of AFM are classified into two categories: D-AFM and AM, and further subdivided into four categories based on the response of the magnon bands under an out-of-plane electric field. The symmetry operations $A$ connecting two magnetic sublattices are divided into four sets I-IV based on whether they cause chiral degeneracy and whether they can be broken by an out-of-plane electric field. The check mark and cross on the right side of the sets indicate whether elements of the sets are allowed or prohibited in the corresponding type, respectively.}
	\label{fig1}
\end{figure*}

Altermagnetism has recently emerged as a distinct magnetic phase that unites the desirable properties of FM and AFM \cite{Smejkal2022, Smejkal2022a, Song2025, Mazin2022, Zhu2024}. Analogous to their electronic counterparts, the magnon bands of altermagnets (AM) exhibit spontaneous chiral splitting in momentum space [Fig.~\ref{fig1}(a)], as recently confirmed in materials such as RuO$_2$, MnTe and $\alpha$-Fe$_2$O$_3$ \cite{Liu2024, Krempasky2024, Fedchenko2024, Smejkal2023, ElKanj2023, Yang2024, Xu2025}. This spontaneous chiral splitting, driven by the non-relativistic exchange anisotropy arising from symmetry breaking, underpins unconventional thermal spin transport phenomena such as the spin Seebeck effect (SSE) and the spin Nernst effect (SNE) \cite{Yang2026, Wu2016, Rezende2016, Kato2025}. However, from an application perspective, the chiral splitting of magnons in AM is challenging to manipulate via external fields, especially electric fields, which restricts its potential in magnon-based spintronic devices.

Electric field control of magnon bands offers an ideal strategy for manipulating spin transport, owing to its low energy consumption, ultrafast response, and ease of integration \cite{Matsukura2015, Kajiwara2010, Xie2025, Wang2024a, Shen2026, Yu2025}. While recent efforts have exploited ferroelectric switching and magnetoelectric coupling in multiferroic materials to modulate spin transport \cite{Huang2024, Parsonnet2022, Chai2024, Wang2025}, the direct and effective manipulation of chiral splitting remains elusive. Specifically, the ability to tune the sign and magnitude of this splitting via electric fields has not yet been demonstrated. Moreover, the absence of a comprehensive symmetry theory describing the electric field response of magnon bands precludes clear theoretical guidance for controlling chiral spin transport by electric fields.

In this work, we present a symmetry classification of the electric field response of magnon bands in two-dimensional (2D) collinear AFM, grounded in our developed theory of collinear spin layer groups (SLGs). We propose the concept of extrinsic chiral splitting, wherein an external electric field lifts the chiral degeneracy, thereby enabling control over both the sign and magnitude of the splitting. We identify the symmetry criteria and the lowest order magnetic exchange paths required to generate alternating magnon bands. Moreover, we demonstrate the electric field control over the direction and magnitude of thermal spin current. This work develops a general theory for the electric-field control of magnon chiral splitting in 2D magnets.

To elucidate the symmetry constraints governing chiral splitting of magnons in 2D magnets, we first constructed 464 collinear SLGs via group extension of the 80 layer groups [Fig.~\ref{fig1}(b)]. Similar to collinear spin space groups \cite{Chen2025, Xiao2024, Chen2024}, the symmetry operations of an SLG are denoted as $[O_{s}||O_{r}]$, where $O_{s}$ and $O_{r}$ represent operations acting on spin space and real space, respectively. Specifically, for 2D FM and FiM, SLGs are described as $[E || \mathbf{G}]$, where $E$ is the identity operation in spin space, and $\mathbf{G}$ is one of the 80 layer groups. For collinear AFM with opposite-spin sublattices, SLGs are described by the extension $[E || \mathbf{H}]+[C_{2}||A\mathbf{H}]$ (total 384), where $[E || \mathbf{H}]$ contains all symmetry operations that do not exchange magnetic sublattices, while $[C_{2}||A\mathbf{H}]$ describes operations that both exchange magnetic sublattices and reverse the spins. 

When the two magnetic sublattices are correlated by the symmetry operator $[C_{2}||P]$ or $[C_{2}||\tau]$, where $P$ and $\tau$ are spatial inversion and lattice translation operations, respectively, the magnon bands keep chiral degeneracy throughout the momentum space [Fig.~\ref{fig1}(a)], similar to three-dimensional magnets. Differently, for 2D magnets, two additional operations $[C_{2}||C_{2z}]$ and $[C_{2}||M_{z}]$, also lead to chiral degeneracy of the magnon bands. In addition, some higher-order rotation operators that exchange the magnetic sublattices, involving $A=\bar{C}_{3z}, C_{6z}, \bar{C}_{6z}$, include $A=P,C_{2z},M_{z}$, respectively, and thus result in chiral degeneracy. The symmetry operator $[C_{2}||C_{3z}]$ is excluded since this work focuses on collinear AFM. When the two sublattices are correlated by other operations $A\in \{C_{4z}, {\bar{C}_{4z}}, C_{2x}, M_{z}\}$, the magnon bands exhibit spontaneous alternating chiral splitting, similar to that of the electronic bands of AM. Therefore, the 384 SLGs of the conpensated AFM can be divided into two categories: one with chiral degeneracy of the magnon bands (292), while the other with alternating chiral splitting (92) [Fig.~\ref{fig1}(b)].  

Applying an electric field can break the symmetry operators that protect the chiral degeneracy, thereby inducing chiral splitting of magnons. The symmetry operations of the SLGs can be classified into four categories based on whether they enforce chiral degeneracy and whether they can be broken by an out-of-plane electric field $E_{z}$, as depicted in Fig.~\ref{fig1}(b). Specifically, the type I involves $A\in \{C_{2z}, C_{6z}, \tau \}$, enforces chiral degeneracy of magnons and remains unbroken by $E_{z}$; the type II involves $A\in \{ P, M_{z}, {\bar{C}_{3z}}, {\bar{C}_{6z}} \}$, leads to chiral degeneracy but is broken by $E_{z}$; the type III involves $A\in \{ C_{4z}, M_{x} \}$, leads to alternating chiral splitting of magnons and remains unbroken by $E_{z}$; while the type IV involves $A\in \{{\bar{C}_{4z}}, C_{2x} \}$, leads to alternating chiral splitting but can be broken by $E_{z}$. Therefore, the 384 SLGs can be further subdivided into four categories based on the response of magnon bands under an out-of-plane electric field. The 200 SLGs that satisfy condition $\exists{A} \in $ {I}$ $ retain chiral degeneracy under $E_{z}$. The 46 SLGs that satisfy condition $\forall{A} \notin $ {I}$ $, $\exists{A} \in $ {II}$ $ and $\exists{A} \in $ {III}$ $ exhibit alternating chiral splitting induced by electric fields, termed extrinsic AM (E-AM). The 63 SLGs that satisfy condition $\forall {A} \notin $ {I} $ \cup $ {III}$ $ exhibit FiM-like chiral splitting  under $E_{z}$, termed E-FiM, while the 75 SLGs that satisfy condition $\forall{A} \notin $ {I} $ \cup $ {II}$ $, $\exists {A} \in $ {III}$ $ retain alternating chiral splitting under $E_{z}$.

Next, the symmetry constrains on the chiral splitting of the magnon bands by the SLGs are revealed based on the linear spin wave theory. The spin Hamiltonian that incorporates various inter- and intra-sublattice magnetic exchange interactions can be described as 
\begin{equation}
\begin{aligned}
\hat{H}=&J_{AB}(\mathbf{R}) \sum_{\langle i, j\rangle} \hat{S}_{Ai} \hat{S}_{Bj} +J_{A}(\mathbf{R}) \sum_{\langle\langle i, j\rangle\rangle} \hat{S}_{Ai} \hat{S}_{Aj} \\& +J_{B}(\mathbf{R}) \sum_{\langle\langle i, j\rangle\rangle} \hat{S}_{Bi} \hat{S}_{Bj}
\end{aligned}
\end{equation}
where the subscript $i,j$ represent the spin sites, $\hat{S}_A$ and $\hat{S}_B$ are the spin operators for sublattices $A$ and $B$, respectively. $J_{AB}$ and $J_{A}/J_{B}$ represent the inter- and intra-sublattice magnetic exchange parameters, respectively, where $\mathbf{R}$ is the connecting vector of spin pairs.

\begin{figure}
	\centering
	\includegraphics[width=1\columnwidth]{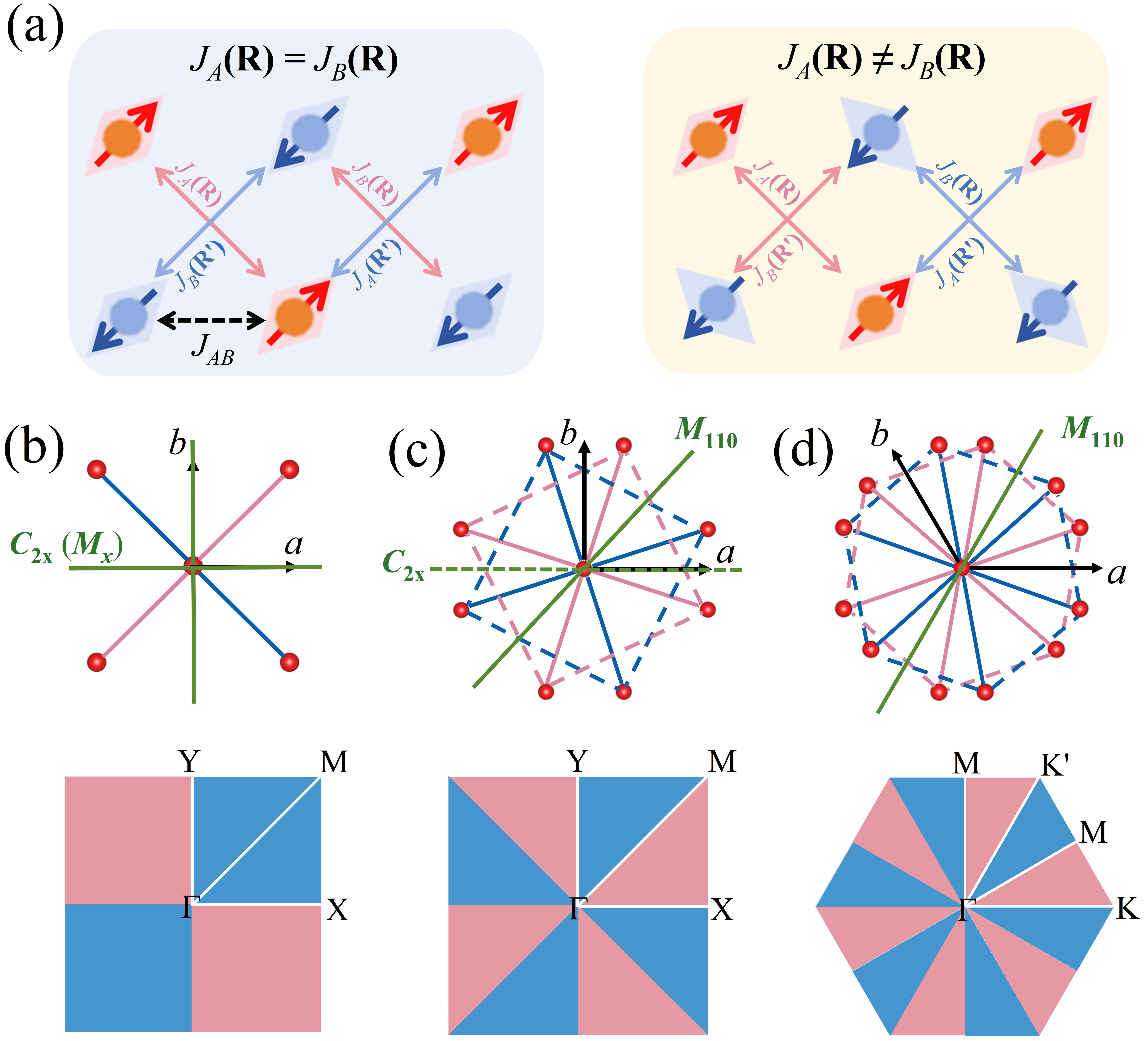}
	\caption{The lowest-order magnetic exchange interaction required for altermagnetic chiral splitting. (a) Schematic of degeneracy and non-degeneracy of intra-sublattice magnetic exchange interactions. Degenerate magnetic exchange interactions are represented by double arrows of the same color. The lowest order magnetic exchange paths determined by SLG operators and the corresponding characteristics of chiral splitting, involving (b) $d$-wave, (c) $g$-wave, and (d) $i$-wave. The blue and red areas represent the opposite signs for chiral splitting.}
	\label{fig2}
\end{figure}

Within the framework of linear spin wave theory, we mapped the spin operators to bosonic operators via the Holstein-Primakoff transformation \cite{Holstein1940}, and then calculated the band dispersion for the two chiral magnon modes through Fourier and Bogoliubov transformations  (see Supplementary Material \cite{SM}). The chiral splitting of magnons can be expressed as:
\begin{equation}
\Delta \omega(\mathbf{k}) \propto \sum_{\mathbf{R}}\left[J_{A}(\mathbf{R})-J_{B}(\mathbf{R})\right][1-\cos (\mathbf{k} \cdot \mathbf{R})].
\end{equation}
Consequently, the existence of inequivalence between the two intra-sublattice magnetic exchange interactions $J_A(\mathbf{R}) \neq J_B(\mathbf{R})$ is crucial for inducing chiral splitting of magnons, as illustrated in Fig.~\ref{fig2}(a). The magnitude of the chiral splitting is proportional to the difference between the two intra-sublattice exchange interactions, while the momentum-space profile of the chiral splitting is governed by the structure factor $1-\cos (\mathbf{k} \cdot \mathbf{R})$.

If the two sublattices are not correlated by any symmetry operation, the magnon bands exhibit chiral splitting throughout the Brillouin zone except at the $\Gamma$ point, similar to the case in FiM. If $\exists {A} \in $ {I} $ \cup $ {II}$ $, then the condition $J_{A}(\mathbf{R}) = J_{B}(\mathbf{R})$ holds for any exchange path, resulting in persistent chiral degeneracy. When $\exists {A} \in $ {III}$ \cup ${IV}$ $, alternating chiral splitting in momentum space is allowed, as shown in Fig.~\ref{fig1}(a). However, for some exchange paths that satisfy $A \mathbf{R}= \pm \mathbf{R}$, the condition $J_{A}(\mathbf{R}) = J_{B}(\mathbf{R})$ is enforced, which means that certain magnetic exchange interactions do not contribute to the chiral splitting. Therefore, alternating chiral splitting necessitates that the spin Hamiltonian includes non-degenerate magnetic exchange interactions.

Taking a 2D square lattice as an example, we illustrate how the SLGs operator determines the lowest-order magnetic exchange interactions required for alternating chiral splitting and the splitting characteristics ($d$-wave or $g$-wave). If $A = C_{4z}$ or ${\bar{C}_{4z}}$, there is no constraint on the magnetic exchange interactions for inducing chiral splitting. In contrast, for $A = C_{2x}$ (or $M_{x}$), magnetic exchange interactions along the ${\langle 100\rangle}$ directions, including the nearest-neighbor ones, do not contribute to chiral splitting. Consequently, next-nearest-neighbor exchange interactions (along the ${\langle 110\rangle}$ directions) or other longer-range exchange must be involved to generate chiral splitting. In momentum space, this symmetry leads to chiral degeneracy along the $\Gamma$–X and $\Gamma$–Y directions, thereby exhibiting a $d$-wave characteristic (see Fig.~\ref{fig2}(b) and Fig. S4 \cite{SM}). When these two symmetry operations exist simultaneously and along different directions, such as $A$ = $C_{2x}$ and $M_{110}$, then the magnetic exchange interactions along these high-symmetry directions are degenerate for the two sublattices. Therefore, the lowest-order magnetic exchange interaction that can cause chiral splitting shifts to the third-nearest-neighbor path (along the ${\langle210\rangle}$ direction) [see Fig.~\ref{fig2}(c)]. This symmetry enforces chiral degeneracy along the $\Gamma$–X, $\Gamma$–M and $\Gamma$–Y  directions in momentum space, manifesting a $g$-wave characteristic(Fig. S4 \cite{SM}). For a 2D hexagonal lattice, only $A$ = $C_{2x}$ (or $M_{x}$) is allowed for alternating chiral splitting, which means that the magnetic exchange interactions are degenerate along directions parallel and perpendicular to the basis vectors. Consequently, the lowest-order magnetic exchange path capable of inducing chiral splitting is along the  ${\langle310\rangle}$ direction. Chiral degeneracy always occurs along the $\Gamma$–K and $\Gamma$–M paths, manifesting a $i$-wave characteristic (Fig.~\ref{fig2}(d) and Fig. S4 \cite{SM}).

Using the established set of 92 SLGs of AM as the screening criterion, we selected Fe$_2$WS$_4$, MnS$_2$, and Mn$_2$P$_2$S$_3$Se$_3$ monolayers as examples to demonstrate the lowest-order magnetic exchange interaction determined by SLGs. Phonon band structures and first-principles molecular dynamics simulations confirm their dynamical and thermodynamic stability in the ground state (Fig. S1 and S2 \cite{SM}), respectively. First-principles calculations confirm that they exhibit collinear antiferromagnetic structure (Fig. S3 \cite{SM}). The magnetic exchange interactions are calculated using the Green's function method based on magnetic force theory, and the magnon band dispersions are calculated using the linear spin wave theory (see Supplementary Material \cite{SM}).

\begin{figure}
	\centering
	\includegraphics[width=1\columnwidth]{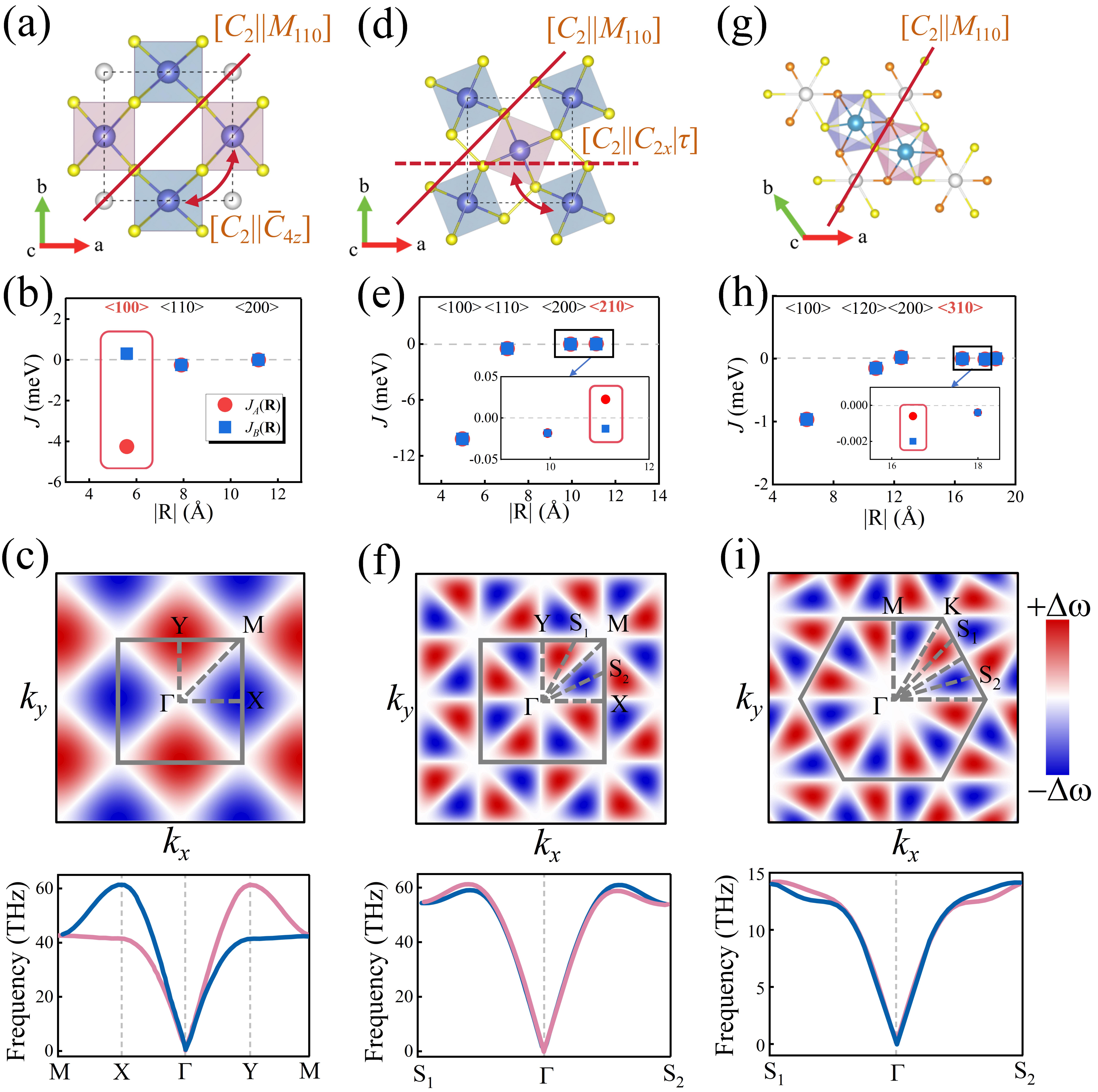}
	\caption{Material candidates with different lowest-order magnetic exchange interactions and chiral splitting types. (a) Crystal structures and symmetry operations connecting two magnetic sublattices, (b) calculated intra-sublattice magnetic exchange interactions, and (c) 2D isosurfaces of chiral splitting and  magnon band dispersions for Fe$_2$WS$_4$ monolayer ($d$-wave). The values of the lowest-order magnetic exchange interactions are circled in a red box, with direction marked in red. (d)-(f) and (g)-(i) show the corresponding results for MnS$_2$ ($g$-wave) and Mn$_2$P$_2$S$_3$Se$_3$ ($i$-wave), respectively.}
	\label{fig3}
\end{figure}

For Fe$_2$WS$_4$ monolayer with a SLG of ${P}^{\overline{\text{1}}} {\overline{\text{4}}}^{\text{1}}\text{2}^{\overline{\text{1}}}{m}$, the two antiferromagnetic sublattices are correlated by $[C_{2}||{\bar{C}_{4z}}]$ and $[C_{2}||M_{110}]$ [Fig.~\ref{fig3}(a)]. The former imposes no restriction on exchange paths for inducing chiral splitting, while the latter causes exchange interactions along the ${\langle 110\rangle}$ directions to remain degenerate. Therefore, the intra-sublattice nearest-neighbor exchange interaction (along ${\langle 100\rangle}$ directions) serves as the lowest-order one capable of inducing chiral splitting [Fig.~\ref{fig3}(b)]. This SLG gives rise to $d$-wave type of magnon bands, characterized by chiral degeneracy along the $\Gamma$–M path while splitting along other paths [Fig.~\ref{fig3}(c)]. For another tetragonal lattice material, MnS$_2$ (${P}^{\text{1}}{\overline{\text{4}}}^{\overline{\text{1}}}\text{2}_{\text{1}}^{\overline{\text{1}}}{m}$), the operations connecting opposite magnetic sublattices are $[C_{2}||M_{110}]$ and $[C_{2}||C_{2x}|\tau]$ [Fig.~\ref{fig3}(d)], which results in the magnetic exchange interactions along the high-symmetry directions not contributing to chiral splitting. Therefore, the lowest-order magnetic exchange interaction shifts to the fourth nearest neighbor, \emph{i.e.}, along the ${\langle210\rangle}$ directions [Fig.~\ref{fig3}(e)]. Similar to magnetic exchange interactions, the magnon bands retain chiral degeneracy along these high-symmetry directions, thus exhibiting a $g$-wave type [Fig.~\ref{fig3}(f)]. For Mn$_2$P$_2$S$_3$Se$_3$ (${P}^{\text{1}}\text{3}^{\text{1}}\text{1}^{\overline{\text{1}}}{m}$) with a hexagonal lattice, only the operator $[C_{2}||M_{110}]$ correlates the two magnetic sublattices [Fig.~\ref{fig3}(g)], resulting in the lowest-order magnetic exchange interaction along the ${\langle310\rangle}$ directions [Fig.~\ref{fig3}(h)]. The magnon bands exhibit a $i$-wave characteristic [Fig.~\ref{fig3}(i)]. Note that this characteristic applies to all SLGs of AM with a 2D hexagonal lattice. 

\begin{figure}[h]
    \centering
	\includegraphics[width=1\columnwidth]{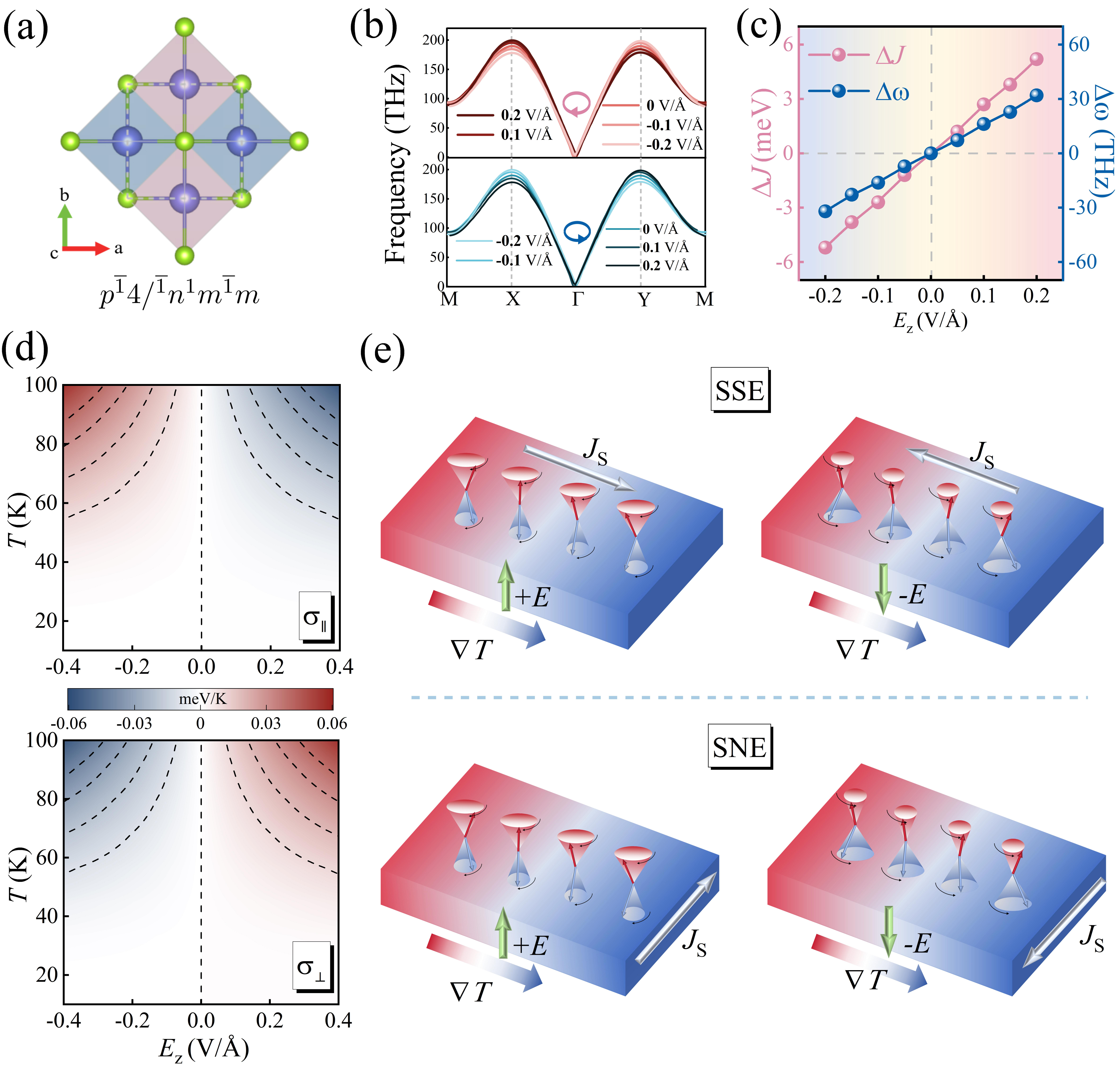}
	\caption{Electric-field control of chiral splitting and spin transport. (a) Crystal structure and SLG of FeSe monolayer. (b) Magnon band structures under different out-of-plane electric field values. (c) Dependence of the anisotropy of the intra-sublattice nearest-neighbor exchange interaction ($\Delta J$) and the maximum chiral splitting ($\Delta\omega$) on the electric field. (d) Calculated coefficients of SSE ($\sigma_{\parallel}$) and SNE ($\sigma_{\perp}$) as functions of temperature and electric field. (e) Schematic illustration of electric field control over thermal spin transport. Changing the sign of the electric field can reverse the spin current caused by SSE and SNE.}
	\label{fig4}
\end{figure}

Then, we take FeSe monolayer as an example to demonstrate the control effect of the electric field on the chiral splitting of an E-AM. Its SLG ${P}^{\overline{\text{1}}} \text{4}/^{\overline{\text{1}}} \mathrm{n}^{\text{1}} \mathrm{m}^{\overline{\text{1}}} \mathrm{m}$ enforces magnon chiral degeneracy throughout the Brillouin zone 
[Fig.~\ref{fig4}(a)]. The magnetic ground state of the system remains unchanged within the range of the applied electric-field strength (Fig. S3 \cite{SM}). When an out-of-plane electric field is applied, all symmetry operators that protect chiral degeneracy are broken, while the operators $[C_{2}||C_{4z}]$ and $[C_{2}||M_{110}]$ are retained. This reduction in symmetry leads to a $d$-wave splitting characteristic, with the maximum splitting occurring at the high-symmetry $X$ and $Y$ points [Fig.~\ref{fig4}(b)]. The chiral splitting primarily stems from the anisotropy of the intra-sublattice nearest-neighbor exchange interaction caused by the electric field. It varies linearly with the electric field strength, and changes sign upon reversal of the electric field [Fig.~\ref{fig4}(c)].

We further investigate the control of spin transport based on magnon chirality by the electric field. The SSE and SNE, as two types of thermal spin transport effects of magnons, both rely on the directional flow of distinct chiral magnon modes ($\alpha$ and $\beta$) driven by temperature gradients \cite{Cui2023, Mook2019}. Within the linear response framework, the longitudinal and transverse spin current density under temperature gradient $\nabla{T}$ can be expressed as 
\begin{equation}
\binom{j_{\|}^{z}}{j_{\perp}^{z}}=\left(\begin{array}{cc}\sigma_{\|} & 0 \\ 0 & \sigma_{\perp}\end{array}\right)\binom{-\partial_{n} T}{-\partial_{n} T}.
\end{equation}
The magnon thermal conductivity is contributed by the $\alpha$ and $\beta$ modes $\sigma_{\parallel} = \sigma_{\parallel}^{\alpha} + \sigma_{\parallel}^{\beta}$ and $\sigma_{\bot} = \sigma_{\bot}^{\beta} - \sigma_{\bot}^{\alpha}$. 
The magnon thermal conductivity caused by each mode depends on the direction of the temperature gradient $\sigma_{\parallel}^{\alpha} = \sigma_{xx}^{\alpha}\cos^{2}\theta + \sigma_{yy}^{\alpha}\sin^{2}\theta$, $\ \sigma_{\bot}^{\alpha} = \left( \sigma_{xx}^{\alpha} - \sigma_{yy}^{\alpha} \right)\cos\theta\sin\theta$, and similarly for $\sigma_{\parallel}^{\beta}$ and $\sigma_{\bot}^{\beta}$, where $\theta$ is the angle between the temperature gradient and the $x$ axis (see Supplementary Material \cite{SM} for detailed derivation). The tensor elements of magnon thermal conductivity can be calculated based on the Kubo formula \cite{Kubo1957, Mook2019, Cui2023}. 

Figure \ref{fig4}(d) shows the variation of the longitudinal and transverse spin thermal conductivity with temperature and electric field. Both increase significantly with increasing temperature, due to the rapid increase in the number of magnons participating in spin transport. The spin conductivities contributed by the two chiral magnon modes have opposite signs, and their magnitudes depend on the frequency of the magnon. This results in the cancellation of the spin conductivities of the two modes in the absence of an electric field. As the electric field increases, the difference in spin conductivity between the two magnon modes enlarges, leading to a gradual increase in the total spin conductivity [Fig.~\ref{fig4}(d)]. Importantly, the direction of the longitudinal and transverse spin currents can be reversed by changing the direction of the electric field. Therefore, based on E-AM, the electric field can not only modulate the strength of thermal spin transport, but also switch of the direction of spin current, as illustrated in Fig.~\ref{fig4}(e).

In conclusion, we have established SLGs framework to classify magnon chiral splitting in 2D magnets. We identify two types of magnets exhibiting extrinsic chiral splitting, termed E-AM and E-FiM, in which the symmetry operators protecting chiral degeneracy can be broken by an electric field. Analogous to the formation of degenerate nodal lines in momentum space, symmetry operators enforce the degeneracy of specific intra-sublattice magnetic exchange interactions, thus determining the lowest-order magnetic exchange interaction required for alternating chiral splitting. We demonstrate precise electric-field control over both the magnitude and sign of the chiral splitting, thereby enabling the reversible switching of the SSE and SNE. This work establishes a design principle for electric-field control over spin transport based on magnon chirality, and offers a promising route for developing magnonic devices.

\begin{acknowledgments} 
\textit{Acknowledgments}---This work was financially supported by the National Natural Science Foundation of China (Grants No. 12374097, No. 11974418 and No. 124B2066) and the Natural Science Foundation of Jiangsu Province (No. BK20233001).
\end{acknowledgments}
\bibliography{reference}
\end{document}